\documentclass[twocolumn,showpacs,preprintnumbers,amsmath,amssymb,superscriptaddress]{revtex4-1}
\usepackage{graphicx}
\graphicspath{ {figs/} }

\usepackage[rflt]{floatflt}
\usepackage{float}

\newcommand{\gsim}{\;\rlap{\lower 3.5 pt \hbox{$\mathchar \sim$}} \raise 1pt
 \hbox {$>$}\;}
\newcommand{\lsim}{\;\rlap{\lower 3.5 pt \hbox{$\mathchar \sim$}} \raise 1pt
 \hbox {$<$}\;}
\newcommand{\GeV}{{\rm GeV}}

\usepackage{color}

\allowdisplaybreaks[4]

\begin{document}

\title{
\boldmath
The effects of  $O(\alpha^2)$ initial state QED corrections to $e^+e^- \rightarrow \gamma^*/Z^*$ 
at very high luminosity colliders}

\author{J. Bl\"umlein}
\affiliation{Deutsches Elektronen--Synchrotron, DESY, Platanenallee 6, D--15738 Zeuthen, Germany}

\author{A. De Freitas}
\affiliation{Deutsches Elektronen--Synchrotron, DESY, Platanenallee 6, D--15738 Zeuthen, Germany}

\author{C.G. Raab}
\affiliation{Institute of Algebra, Johannes Kepler University, Altenbergerstra\ss{}e 69, A--4040, Linz, Austria}

\author{K. Sch\"onwald}
\affiliation{Deutsches Elektronen--Synchrotron, DESY, Platanenallee 6, D--15738 Zeuthen, Germany}

\date{30.09.2019}

\begin{abstract}
\noindent 
We present numerical results on the recently completed $O(\alpha^2)$ initial state corrections to the process
$e^+e^- \rightarrow \gamma^*/Z^*$, which is a central process at past and future high energy and high luminsoity 
colliders for precision measurements of the properties of the $Z$-boson, the Higgs boson, and the top quark.
We observe differences to an earlier result \cite{Berends:1987ab} in the non-logarithmic contributions at $O(\alpha^2)$. 
The new result leads to a  4 MeV shift in the $Z$ width considering the lower end $s_0 = 4 m_\tau^2$ of the radiation 
region, which is larger than the present accuracy. We present predictions on the radiative corrections to the central 
processes $e^+e^- \rightarrow \gamma^*/Z^*$, $e^+e^- \rightarrow Z H$ and $e^+e^- \rightarrow t \overline{t}$ planned 
at future colliders like the ILC, CLIC, Fcc\_ee and CEPC to measure the mass and the width of the $Z$ boson, the Higgs 
boson and the top quark, for which the present corrections are significant.
\end{abstract}
\preprint{
DESY 19--162, DO--TH 19/18, SAGEX-19-22 
}
 
\maketitle

\noindent 
An important ingredient to precision measurements at $e^+e^-$ colliders is the precise knowledge of the QED initial 
state corrections (ISR). The $O(\alpha^2)$ corrections have been completed very recently. Already in 1987 a first 
calculation to $O(\alpha^2)$ has been performed \cite{Berends:1987ab} for the process $e^+e^- \rightarrow \gamma^*/Z^*$. 
These corrections have been used in the analysis of the LEP1 data, cf.~\cite{ALEPH:2005ab} and are implemented in 
fitting codes like  {\tt TOPAZ0} \cite{Montagna:1998kp} and {\tt ZFITTER} \cite{ZFITTER}. In 2011, using the light cone 
expansion and assuming the factorization of the massive Drell-Yan process, the corrections for the same process have 
been calculated in \cite{Blumlein:2011mi} and disagreement was found with the results of \cite{Berends:1987ab} for the 
non-logarithmic terms at $O(\alpha^2)$. 

We have repeated the calculation using conventional methods without performing any approximation and expanded the final 
results in the mass ratio $m_e^2/s$ to obtain compact analytic expressions for the respective radiators, 
cf.~\cite{Blumlein:2019srk,DYLONG}. The calculation has been accompanied by controlling the results using high precision 
numerics. We confirm the results presented in \cite{Blumlein:2011mi}. Furthermore, in  Ref.~\cite{Berends:1987ab} no 
account was given on the axialvector terms, which have different corrections than the vector terms in some cases. Also some 
processes only 
contributing to the non-logarithmic order known from \cite{Hamberg:1990np,Harlander:2002wh} were missing, which we 
have recalculated and added, completing the $O(\alpha^2)$ QED ISR corrections. Here we include both photon and $e^+e^-$ 
pair emission up to $O(\alpha^2)$. The initial state QED corrections can be written in terms of the following 
functions 
\begin{eqnarray} 
H\left(z,\alpha,\frac{s}{m^2}\right) &=& \delta(1-z) + \sum_{k=1}^\infty \left(\frac{\alpha}{4\pi}\right)^k 
C_k\left(z, \frac{s}{m^2}\right) \\ C_k\left(z, \frac{s}{m^2}\right) &=& \sum_{l=0}^k \ln^{k-l}\left(\frac{s}{m^2}\right) 
c_{k,l}(z), 
\end{eqnarray} 
which yield 
the respective differential cross sections by 
\begin{eqnarray} 
\frac{d \sigma_{e^+e^-}}{ds'} = \frac{1}{s} \sigma_{e^+e^-}(s') H\left(z,\alpha,\frac{s}{m^2}\right), 
\end{eqnarray} 
with $\sigma_{e^+e^-}(s')$ the scattering cross section without the ISR QED corrections, $\alpha \equiv \alpha(s)$ the 
fine structure constant and $z = s'/s$, where $s'$ is the invariant mass of the produced (off-shell) $\gamma/Z$ boson.

These results are of phenomenological importance for the precision measurements of the $Z$ resonance, high luminosity 
$Z H$ production, and $t\overline{t}$ production at LEP1, and for future planned $e^+e^-$ colliders  such as ILC and 
CLIC \cite{ILC}, the FCC\_ee \cite{FCCEE,Abada:2019zxq}, the CEPC \cite{CEPC}, and also for muon colliders 
\cite{Delahaye:2019omf}.

In this letter we detail the phenomenological results for the impact of the ISR QED corrections up to 
$O(\alpha^2)$ and also 
include soft resummation beyond this order, cf. e.g. \cite{Berends:1987ab}, studying their effect on the $Z$ peak, 
$ZH$- and $t\bar{t}$-production. These processes will serve to perform highly precise measurements of the $Z$ and Higgs 
boson, $H$, and the top quark mass in the future. Likewise, we reconsider the measurement at LEP1. A detailed account on the 
analytic calculation will be given in \cite{DYLONG}, providing also all the radiation functions needed in the analyses, 
which are too voluminous to be presented here.

\vspace*{3mm}
\noindent
{\bf 1.} {\it The $Z$ peak and its surrounding.}\\
For this production channel we consider the measurement of the inclusive cross section of a $\mu^+\mu^-$  state above a 
certain threshold $s_0$ of its invariant mass squared, while all the radiation products due to ISR are integrated. The 
theoretical value for $s_0$ is $4 m_\mu^2$, while in the measurements a series of cuts are used and then one extrapolates 
again to a value of $s_0$. In the LEP1 analysis, examples are $s_0 = 4 m_\tau^2$ or $s_0 = 0.01 M_Z^2$ \cite{ALEPH:2005ab}. 
We will discuss effects for these values and also consider values down to the theoretical boundary. 

In Table~\ref{TAB1} we summarize the effect of the different order ISR corrections on the shift of the $Z$ peak and 
the modification of the half-width performing the difference from a given order to the previous one. Very similar 
values are obtained in the case of a fixed width or the $s$-dependent width. At $O(\alpha^2)$ we distinguish the cases of 
either pure
\begin{figure}[t]
  \centering
  \hskip-0.8cm
  \includegraphics[width=.9\linewidth]{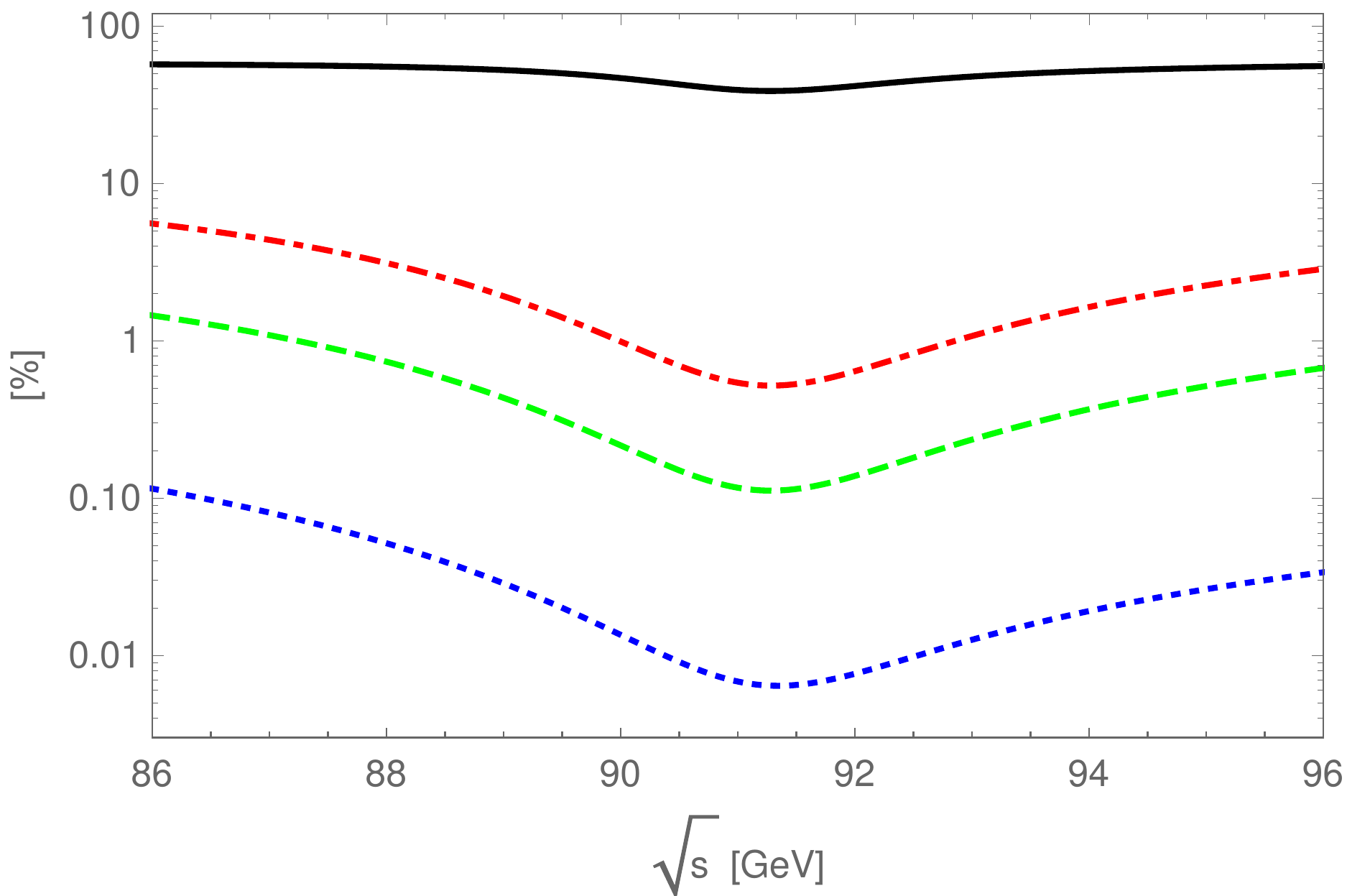}
  \caption[]{\sf Relative difference between the $O(\alpha^2)$ results of \cite{Berends:1987ab} and the present paper as a 
  function of $\sqrt{s}$ in dependence of $s_0$. 
  Dotted line $s_0 = 0.01 M_Z^2$; 
  Dashed line $s_0 = 4 m_\tau^2$; 
  Dash-dotted line $s_0 = 1 \GeV^2$; 
  Full line $s_0 = 4 m_\mu^2$.} 
  \label{fig:cutZ}
\end{figure}

\vspace*{-1cm}
\noindent
photon emission 
or including also $e^+e^-$ pair production. While the peak shift comes out the same in both cases, 
there is a shift on the width of 28 MeV by including the emission of $e^+e^-$ pairs. Finally, soft photon 
exponentiation 
from $O(\alpha^3)$ onward leads to a peak shift of 17 MeV and to a 23 MeV width shift. The numbers are quite comparable 
to those given in \cite{Berends:1987ab}, where at $O(\alpha^2)$ only the photon emission has been considered and the 
integration was performed from $s_0 = 4 m_\mu^2$. 
\begin{table}[H]
\centering
\begin{tabular}{|l|r|r|r|r|}
\hline
\multicolumn{1}{|c|}{} &
\multicolumn{2}{c|}{Fixed width} &
\multicolumn{2}{c|}{$s$ dep. width} \\
\hline
\multicolumn{1}{|c|}{} &
\multicolumn{1}{c|}{Peak} &
\multicolumn{1}{c|}{Width}    &
\multicolumn{1}{c|}{Peak} &
\multicolumn{1}{c|}{Width} \\
\multicolumn{1}{|c|}{} &
\multicolumn{1}{c|}{(MeV)} &
\multicolumn{1}{c|}{(MeV)}    &
\multicolumn{1}{c|}{(MeV} &
\multicolumn{1}{c|}{(MeV)} \\
\hline 
$O(\alpha)$   correction                 &  210 &  603 &  210 &  602 \\
$O(\alpha^2)$ correction                 & -109 & -187 & -109 & -187 \\
$O(\alpha^2)$: $\gamma$ only             & -110 & -215 & -110 & -215 \\
$O(\alpha^2)$ correction                 &      &      &      &      \\
+ soft exp.                              &   17 &   23 &   17 &   23 \\
Difference to $O(\alpha^2)$ \cite{Berends:1987ab}      &      &    4 &      &    4 \\
\hline
\end{tabular}
\caption[]{\sf Shifts in the $Z$-mass and the width due to the different contributions to the ISR QED 
radiative corrections for a fixed width of $\Gamma_Z =  2.4952~\GeV$  and $s$-dependent width using $M_Z = 91.1876~\GeV$  
\cite{PDG} and $s_0 = 4 m_\tau^2$, cf.~\cite{ALEPH:2005ab}.}
\label{TAB1}
\end{table}

\noindent
At $s_0 = 4 m_\tau^2$ the corrected expressions w.r.t. Ref. \cite{Berends:1987ab} are too small to be visible at the peak 
position. However, a 4 MeV shift is obtained in the width, in comparison with the present result. This is of relevance since 
the current error is $\Delta \Gamma_Z = \pm 2.3$ MeV \cite{PDG}. For $s_0 = 0.01 M_Z^2$, on the other hand, the shift 
amounts to 0.2 MeV, which is relevant at  Giga-Z and Fcc\_ee \cite{ILC,Abada:2019zxq}, where resolutions of a few 
hundred  keV can be reached  for both $M_Z$ and $\Gamma_Z$, see also \cite{dEnterria:2016sca}. If $s_0$ would have 
been chosen as low as 1 GeV$^2$, the width would shift by 18 MeV and the peak position by 3 keV, while for larger cuts 
the effect on the peak shift cannot be resolved. The effects would even be larger for $s_0 = 4 m_\mu^2$.
To clarify this further, we show in Figure~1 the relative difference of the correction for a series  of $s_0$ values 
in the vicinity of the $Z$ peak.

The shifts in the width are majorly caused by the discrepancies in the pure singlet terms (process 3 in 
\cite{Berends:1987ab}) containing $1/z$ contributions, cf.~\cite{Blumlein:2019srk}.

\begin{figure}[H]
  \centering
  \hskip-0.8cm
  \includegraphics[width=.9\linewidth]{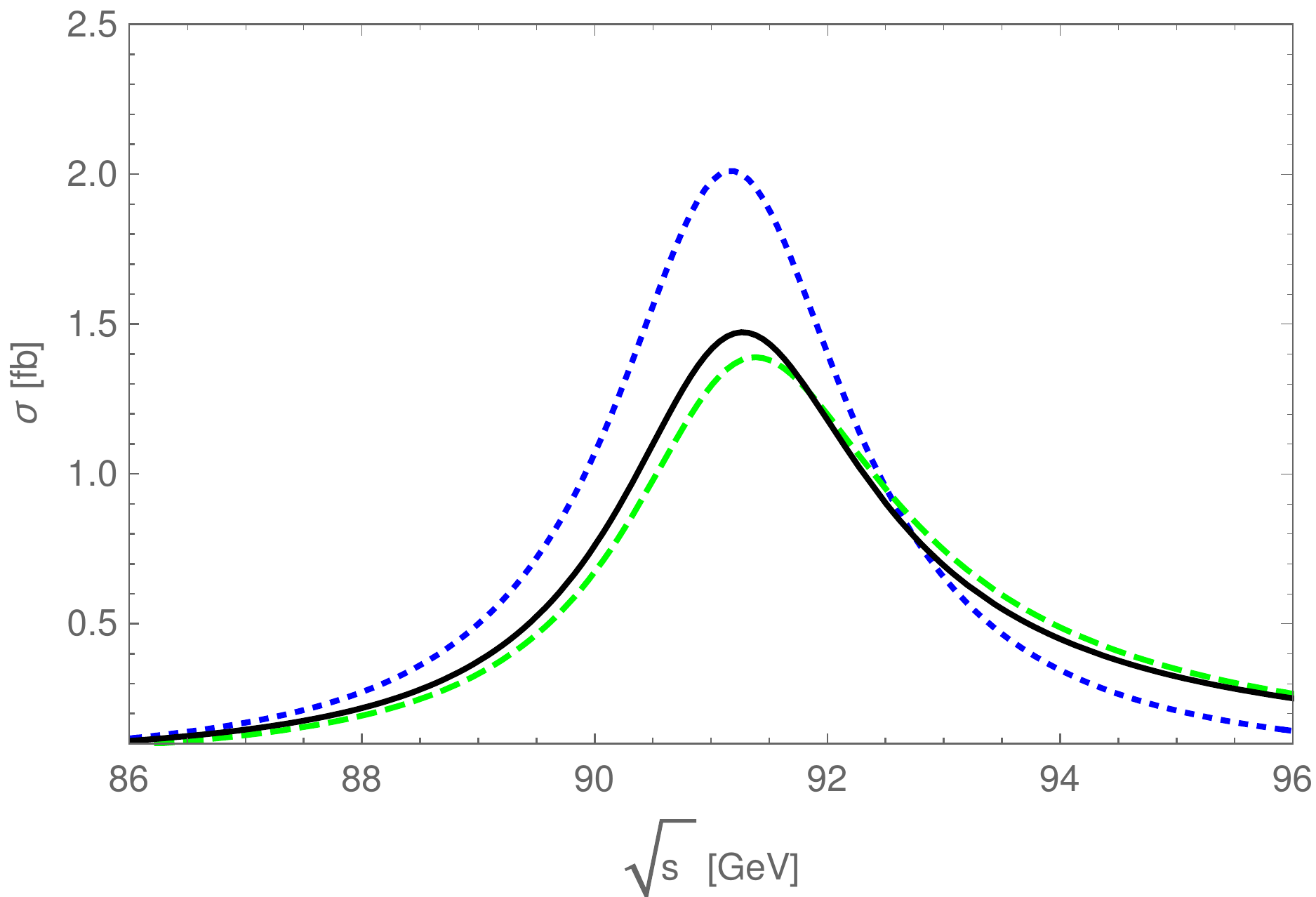}
  \caption[]{\sf The $Z$-resonance in $e^+e^- \rightarrow \mu^+\mu^-$. Dotted line: Born cross section; Dashed line: 
  $O(\alpha)$ ISR corrections; 
  Full line: $O(\alpha^2)$ + soft resummation ISR corrections, with $s_0 = 4 m_\tau^2$.} 
  \label{fig:ZPEAK2}
\end{figure}

\begin{figure}[H]
  \centering
  \hskip-0.8cm
  \includegraphics[width=.9\linewidth]{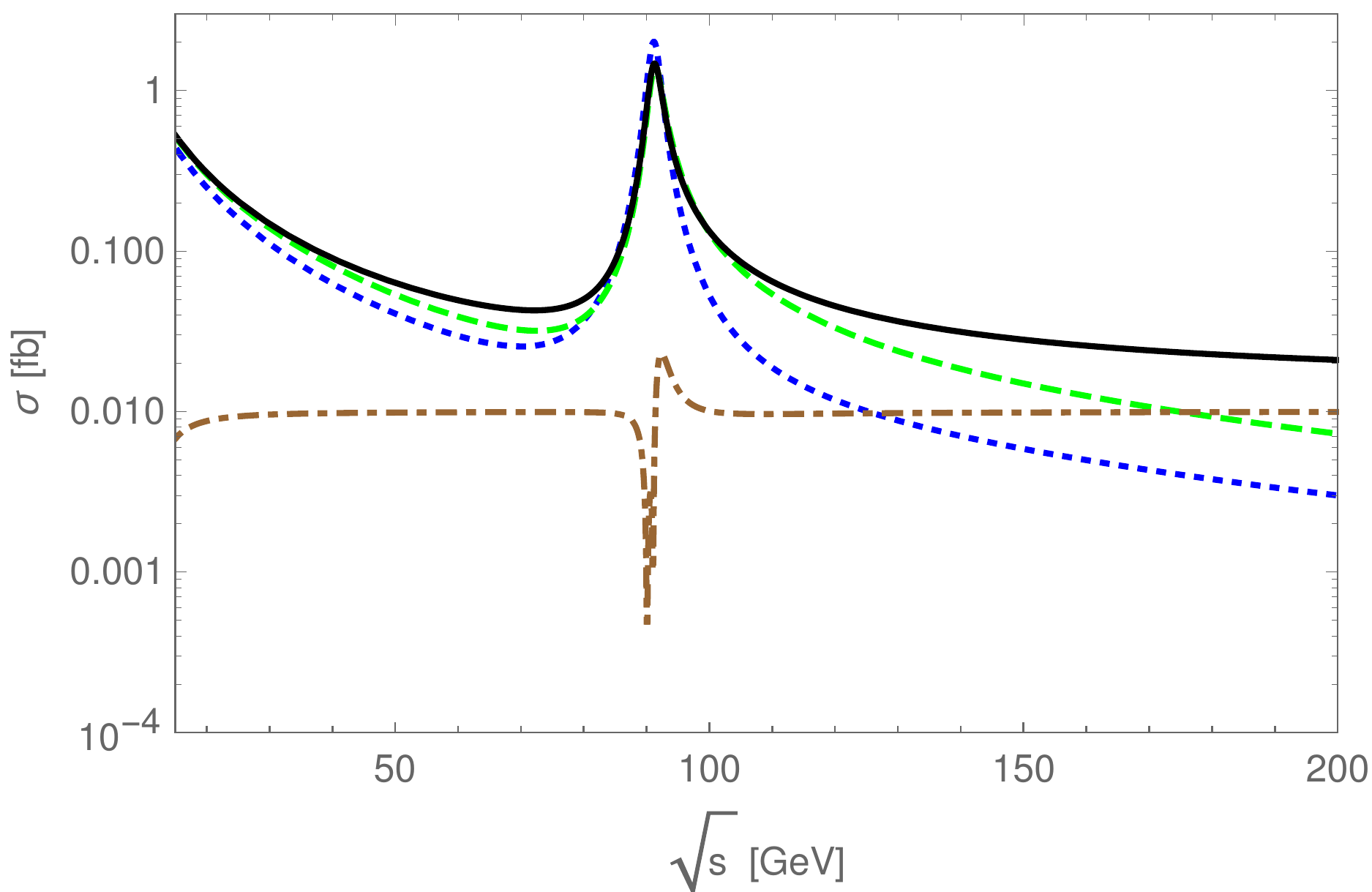}
  \caption[]{\sf The $Z$-resonance in $e^+e^- \rightarrow \mu^+\mu^-$. Dotted line: Born cross section; Dashed line: 
  $O(\alpha)$ ISR corrections; 
  Full line: $O(\alpha^2)$ + soft resummation ISR corrections; Dash-dotted line: individual contribution of soft 
  resummation.} 
  \label{fig:ZPEAK1}
\end{figure}

Between the cases of a constant width and the $s$-dependent width we find a peak shift of $34.2$ MeV and a shift of 
the width of 1 MeV, irrespective of the applied ISR corrections, in accordance with Refs.~\cite{SDEP}.
In Figure~\ref{fig:ZPEAK2} we illustrate the different QED ISR corrections to $e^+e^- \rightarrow Z^*/\gamma^*$ around the 
$Z$ peak. The ISR corrections change the profile of the resonance, i.e. the peak position, height and the half width.
The lines for the $O(\alpha^2)$ correction and the one including soft resummation are nearly identical.
In Figure~\ref{fig:ZPEAK1} the region of $\sqrt{s}$ is extended to $[10,200]~\GeV$. The individual contributions 
of the fixed order corrections at low order show growing effects off the $Z$ peak. The soft resummation corrections 
stay nearly constant in the whole range, except in the region around the $Z$ peak.

\noindent
The full $O(\alpha^2)$ corrections prove to be already important in the analysis of the LEP1. The difference to the 
previous results \cite{Berends:1987ab} has an effect when analyzing the LEP1 data, applying a lower cut of the size 
$s_0 = 4 m_\tau^2$, and likewise $s_0 = 0.01$, for the future measurements at Giga-Z and Fcc\_ee.

\begin{figure}[H]
  \centering
  \hskip-0.8cm
  \includegraphics[width=.9\linewidth]{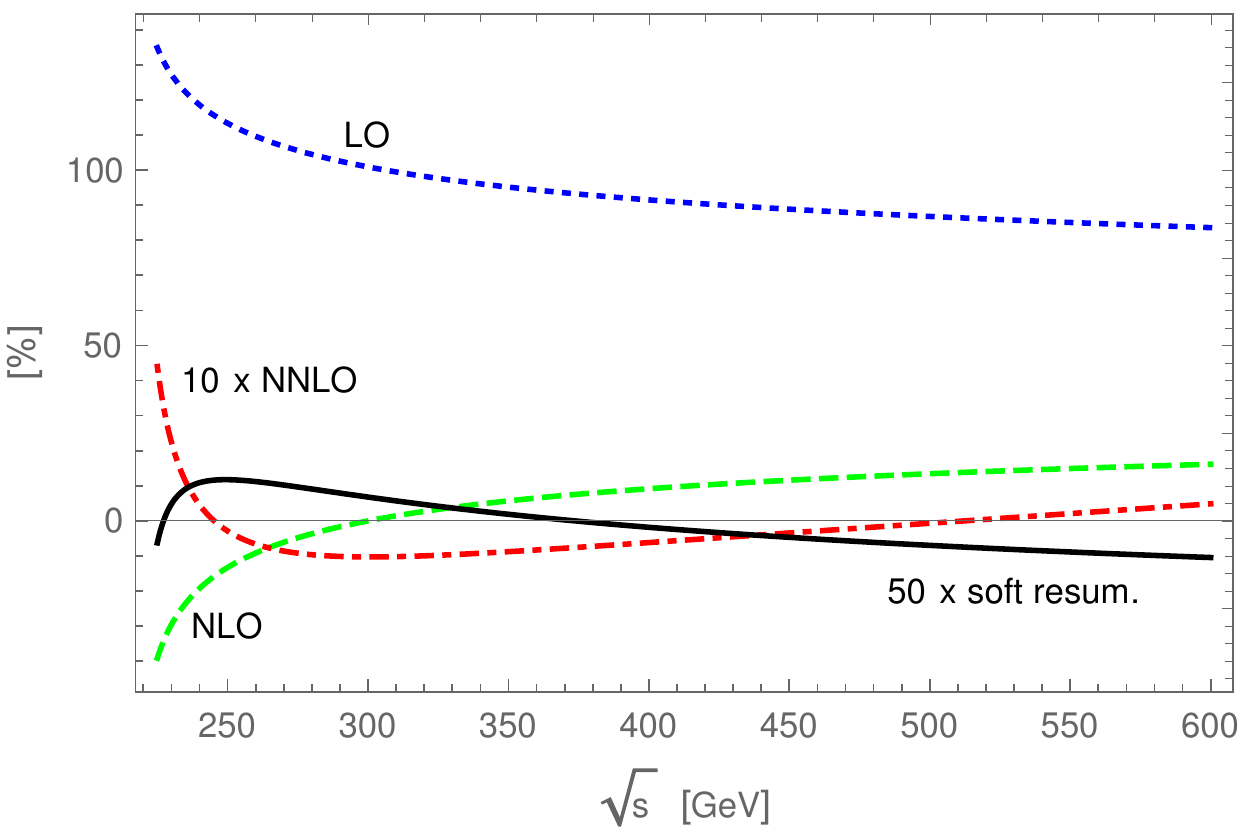}
  \caption[]{\sf Relative contributions of the ISR QED corrections to the cross section for $e^+e^- \rightarrow Z H$ in \%. 
  Dotted line: $O(\alpha^0)$;
  Dashed line: $O(\alpha)$;
  Dash-dotted line: $O(\alpha^2)$;
  Full line: soft resummation beyond $O(\alpha^2)$, with $s_0 = 4 m_\tau^2$.}
  \label{fig:ZH3}
\end{figure}

\vspace*{3mm}
\noindent
{\bf 2.} {\it The process $e^+e^- \rightarrow Z H$.}\\
For the study of the radiative corrections we refer to the Born cross section given in Ref.~\cite{Barger:1993wt}.
The accuracy of the cross section measurement has been estimated to reach 1\% \cite{dEnterria:2016sca} at future colliders like 
the ILC, CLIC, and $0.4\%$ at the Fcc\_ee \cite{Ruan:2014xxa}. In Figure~\ref{fig:ZH3} we show the 
relative
contributions of the Born and the different ISR radiative corrections to $Z H$-production.

\vspace*{-5mm}
\noindent
The NNLO corrections vary between +4.8\% and $-1\%$ and are larger or of the size of the expected
experimental errors. The corrections due to soft resummation are of $O(\pm 0.2\%)$ and reach half of the
projected accuracy. 

\vspace*{3mm}
\noindent
{\bf 3.} {\it The $t\overline{t}$-production at threshold and in the continuum.}

\vspace*{-3mm}
\begin{figure}[H]
  \centering
  \hskip-0.8cm
  \includegraphics[width=.9\linewidth]{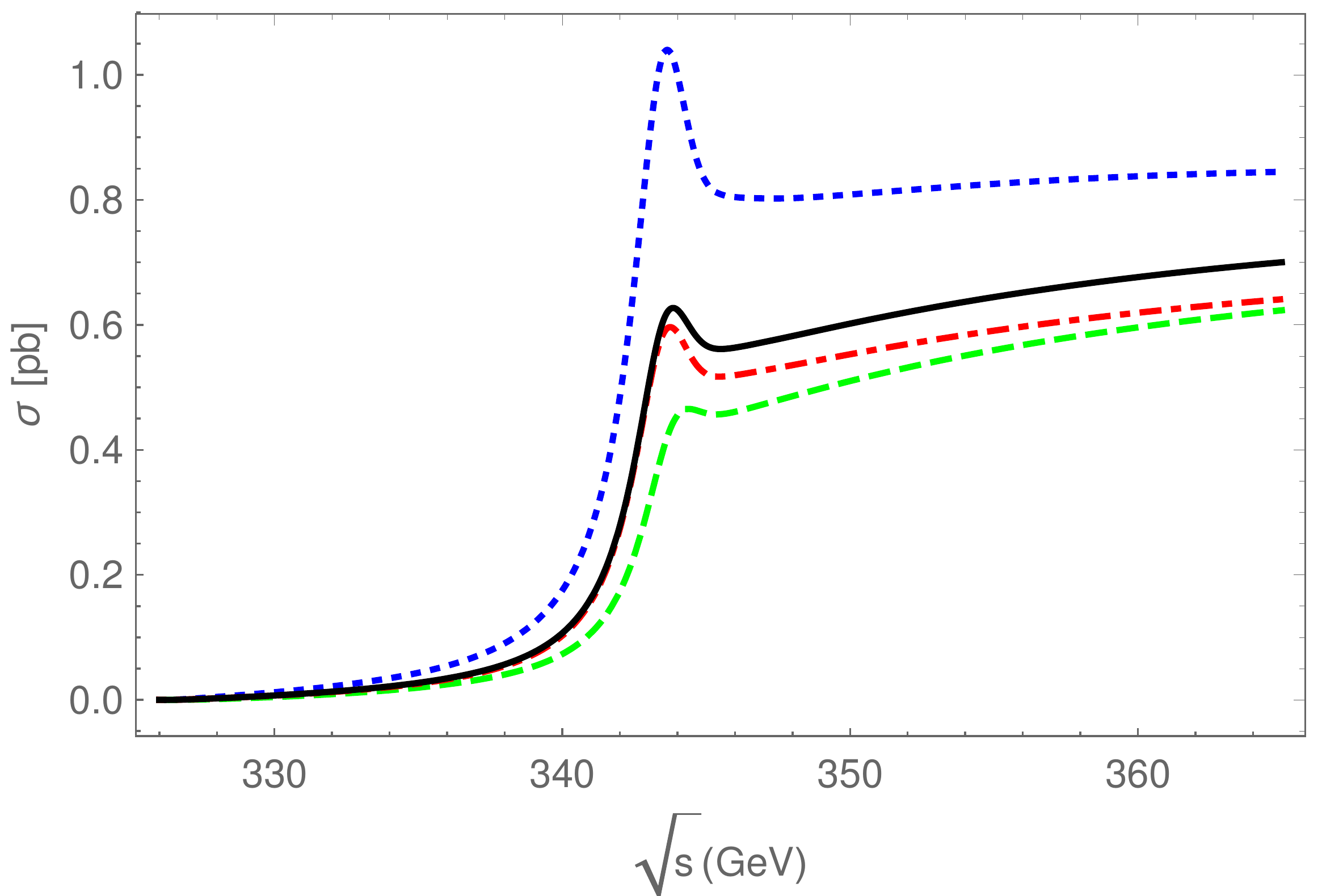}
  \caption[]{\sf The QED  ISR corrections to $e^+e^- \rightarrow t\overline{t}$ ($s$-channel photon exchange) in the threshold 
  region far a PS-mass of $m_t = 172~\GeV$.
Dotted line $O(\alpha^{0})$;
Dashed line $O(\alpha)$;
Dash-dotted line $O(\alpha^{2})$;
Full line $O(\alpha^{2})$ + soft resummation.
  \label{fig:TT1}}
\end{figure}
\noindent
For the process of $e^+e^- \rightarrow t\overline{t}$ we consider the ISR effects both in the threshold 
and the continuum region. In the former case they are applied to the cross section based on including the N$^3$LO
QCD corrections implemented in the code {\tt QQbar\_threshold} \cite{Beneke:2016kkb,Beneke:2017rdn,Beneke:2015kwa}, 
while in the continuum case for $\sqrt{s} > 500 \GeV$ we use the Born cross section \cite{Berends:1987ab} for a first 
numerical illustration. The anticipated accuracy to measure this scattering cross section at future $e^+e^-$ 
colliders has been estimated to be $\pm 2\%$ \cite{Seidel:2013sqa,Simon:2016pwp}.

For the top-quark mass we refer to the PS mass of $172~\GeV$. The corrections in the threshold regions are shown in 
Figure~\ref{fig:TT1}. The different corrections change the profile of the cross section significantly. Up to $\sqrt{s} 
\sim 344~\GeV$ the contributions due to soft resummation agree with the NNLO corrections. Above they deliver an additional
contribution. Adding soft exponentiation implies a correction between 2 and 8\%. Both the $O(\alpha^2)$ and soft 
resummation corrections have effects of the size of the expected experimental accuracy and larger.

\begin{figure}[H]
  \centering
  \hskip-0.8cm
  \includegraphics[width=.9\linewidth]{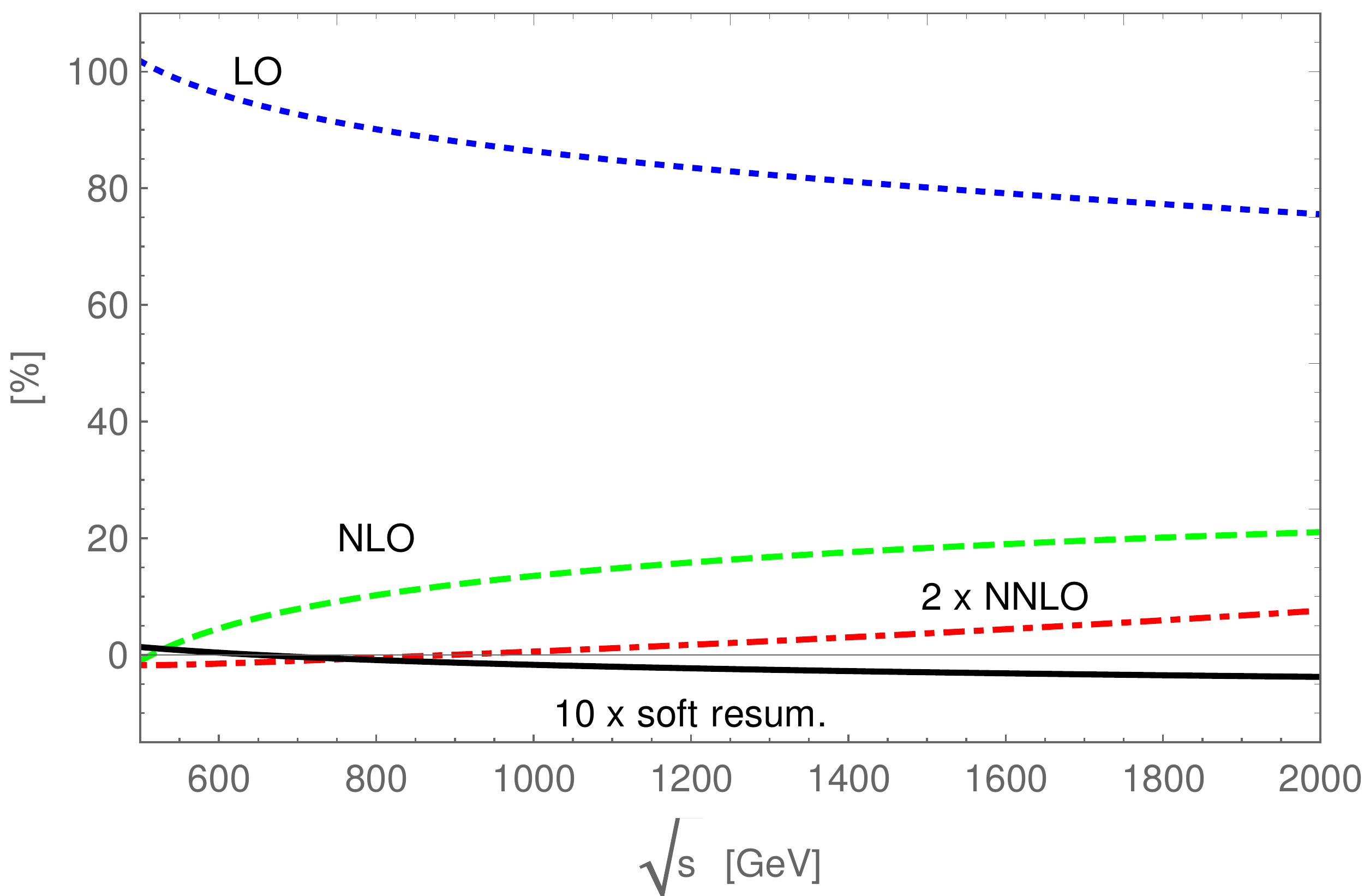}
  \caption[]{\sf Relative contributions of the continuum cross section of $t\overline{t}$ production including the NNLO ISR 
  corrections. 
  Dotted line: $O(\alpha^0)$;
  Dashed line: $O(\alpha)$;
  Dash-dotted line: $O(\alpha^2)$ scaled by 2;
  Full line: soft resummation beyond $O(\alpha^2)$ scaled by 10.}
  \label{fig:TT2}
\end{figure}

In the continuum region the relative size of the ISR corrections to $t\overline{t}$ production are shown in 
Figure~\ref{fig:TT2}. The $O(\alpha^2)$ corrections vary between $-1$ and $4$\% and soft resummation yields further 
corrections of 0.13 to $-0.38\%$.
 
\noindent

\noindent
Whether or not the difference to the results given in \cite{Berends:1987ab} is visible depends on the range in $z$
over which is integrated. Only at small values of $z$ the effect is visible. At higher cuts in $\sqrt{s}$, as the case
for $Z H$- and $t\bar{t}$-production, the numerical effects are very small, given the respective collider energies. 

We designed the {\tt FORTRAN}-code {\tt RC2.f} for the numerical calculations of the ISR corrections [which can be compiled
together with other {\tt FORTRAN}- and {\tt C}-codes by {\tt gfortran} \cite{GF}] for data analyses. We also used an 
implementation in {\tt mathematica}. 

In conclusion, the numerical investigation of the $Z$ boson production, as well as $Z H$ and $t\overline{t}$ production has 
shown 
the relevance of these effects for LEP1 and at future $e^+e^-$ colliders. The new results, compared with  
\cite{Berends:1987ab} imply a relative shift in the $Z$-width by $\sim$4 MeV for $s_0 = 4 m_\tau^2$.
\section*{Acknowledgments}

\noindent 
This paper is dedicated to the memory of our colleague W.L.~van Neerven. We would like to thank M.~Beneke, M.~Gr\"unewald, 
W.~Lohmann, A.~Maier, and P.~Marquard for discussions. This project has received funding from the European Union’s Horizon 
2020 research and innovation programme under the Marie Sk\/{l}odowska-Curie grant agreement No. 764850, SAGEX, and COST action 
CA16201: Unraveling new physics at the LHC through the precision frontier and from the Austrian FWF grants P 27229 and P 31952
in part.


\begin{thebibliography}{99}
%
\bibitem{Berends:1987ab}
  F.A.~Berends, W.L.~van Neerven and G.J.H.~Burgers,
  Nucl.\ Phys.\ B {\bf 297} (1988) 429--478 Erratum: [Nucl.\ Phys.\ B {\bf 304} (1988) 921--922];\\
  B.A.~Kniehl, M.~Krawczyk, J.H.~K\"uhn and R.G.~Stuart,
  Phys.\ Lett.\ B {\bf 209} (1988) 337--342.
%
\bibitem{ALEPH:2005ab}
  S.~Schael {\it et al.} 
  Phys.\ Rept.\  {\bf 427} (2006) 257--454
  [hep-ex/0509008];\\
  J.~Mnich,
  Phys.\ Rept.\  {\bf 271} (1996) 181--266.
%
\bibitem{Montagna:1998kp}
  G.~Montagna, O.~Nicrosini, F.~Piccinini and G.~Passarino,
  Comput.\ Phys.\ Commun.\  {\bf 117} (1999) 278--289
  [hep-ph/9804211];\\
  D.Y.~Bardin and G.~Passarino,
  {\sf The standard model in the making: Precision study of the electroweak interactions},
  International series of monographs on physics. {\bf 104} (Calendron Press, Oxford, 1999).
%
\bibitem{ZFITTER}
  D.Y.~Bardin {\it et al.}, 
  Comput.\ Phys.\ Commun.\  {\bf 133} (2001) 229--395
  [hep-ph/9908433];\\
  A.B.~Arbuzov {\it et al.}, 
  Comput.\ Phys.\ Commun.\  {\bf 174} (2006) 728--758
  [hep-ph/0507146].
%
\bibitem{Blumlein:2011mi}
  J.~Bl\"umlein, A.~De Freitas and W.L.~van Neerven,
  Nucl.\ Phys.\ B {\bf 855} (2012) 508--569
  [arXiv:1107.4638 [hep-ph]].
%
\bibitem{Blumlein:2019srk}
  J.~Bl\"umlein, A.~De Freitas, C.G.~Raab and K.~Sch\"onwald,
  Phys.\ Lett.\ B {\bf 791} (2019) 206--209
  [arXiv:1901.08018 [hep-ph]].
%
\bibitem{DYLONG}
  J.~Bl\"umlein, A.~De Freitas, C.G.~Raab and K.~Sch\"onwald, DESY 18--196.
%
\bibitem{Hamberg:1990np}
  R.~Hamberg, W.L.~van Neerven and T.~Matsuura,
  Nucl.\ Phys.\ B {\bf 359} (1991) 343--405
   Erratum: [Nucl.\ Phys.\ B {\bf 644} (2002) 403--404].
%
\bibitem{Harlander:2002wh}
  R.V.~Harlander and W.B.~Kilgore,
  Phys.\ Rev.\ Lett.\  {\bf 88} (2002) 201801
  [hep-ph/0201206].
%
\bibitem{ILC}
  E.~Accomando {\it et al.} 
  Phys.\ Rept.\  {\bf 299} (1998) 1--78
  [hep-ph/9705442];\\
  J.A.~Aguilar-Saavedra {\it et al.} 
  hep-ph/0106315;\\
{\tt http://www.linearcollider.org/ILC}
\\ 
R.~Franceschini {\it et al.}, arXiv:1812.07986 [hep-ex].
%
\bibitem{Abada:2019zxq}
  A.~Abada {\it et al.} [FCC Collaboration],
  Eur.\ Phys.\ J.\ ST {\bf 228} (2019) no.2,  261--623.
%
\bibitem{FCCEE}
{\tt http://tlep.web.cern.ch/}
%
\bibitem{CEPC}
{\tt http://cepc.ihep.ac.cn/} 
%
\bibitem{Delahaye:2019omf}
  J.P.~Delahaye {\it et al.},
  {\it Muon Colliders},
  arXiv:1901.06150 [physics.acc-ph].
%
\bibitem{PDG}
M. Tanabashi et al. (Particle Data Group), Phys. Rev. D {\bf 98} (2018) 030001 and 2019 update. 
%
\bibitem{dEnterria:2016sca}
  D.~d'Enterria, in:
  {\sf Particle Physics at the Year of Light}, (World Scientific, Singapore, 2017) 182--191, ed. A.I. Studentkin
  arXiv:1602.05043 [hep-ex]; Slides: Higgs Couplings '17, Heidelberg Nov. 10, 2017.
%
\bibitem{SDEP}
  F.A.~Berends, G.~Burgers, W.~Hollik and W.L.~van Neerven,
  Phys.\ Lett.\ B {\bf 203} (1988) 177--182;\\
  D.Y.~Bardin, A.~Leike, T.~Riemann and M.~Sachwitz,
  Phys.\ Lett.\ B {\bf 206} (1988) 539--542;\\
  W.~Beenakker and W.~Hollik,
  Z.\ Phys.\ C {\bf 40} (1988) 141--148.
%
\bibitem{Barger:1993wt}
  V.D.~Barger, K.M.~Cheung, A.~Djouadi, B.A.~Kniehl and P.M.~Zerwas,
  Phys.\ Rev.\ D {\bf 49} (1994) 79--90
  [hep-ph/9306270].
%
\bibitem{Ruan:2014xxa}
  M.~Ruan,
  Nucl.\ Part.\ Phys.\ Proc.\  {\bf 273-275} (2016) 857--862
  [arXiv:1411.5606 [hep-ex]].
%
\bibitem{Beneke:2016kkb}
  M.~Beneke, Y.~Kiyo, A.~Maier and J.~Piclum,
  Comput.\ Phys.\ Commun.\  {\bf 209} (2016) 96--115
  [arXiv:1605.03010 [hep-ph]].
%
\bibitem{Beneke:2017rdn}
  M.~Beneke, A.~Maier, T.~Rauh and P.~Ruiz-Femenia,
  JHEP {\bf 1802} (2018) 125
  [arXiv:1711.10429 [hep-ph]].
%
\bibitem{Beneke:2015kwa}
  M.~Beneke, Y.~Kiyo, P.~Marquard, A.~Penin, J.~Piclum and M.~Steinhauser,
  Phys.\ Rev.\ Lett.\  {\bf 115} (2015) no.19,  192001
  [arXiv:1506.06864 [hep-ph]].
%
\bibitem{Seidel:2013sqa}
  K.~Seidel, F.~Simon, M.~Tesar and S.~Poss,
  Eur.\ Phys.\ J.\ C {\bf 73} (2013) no.8,  2530
  [arXiv:1303.3758 [hep-ex]].
%
\bibitem{Simon:2016pwp}
  F.~Simon,
  PoS (ICHEP2016) 872
  [arXiv:1611.03399 [hep-ex]].
%
\bibitem{GF}
{\tt https://gcc.gnu.org/wiki/GFortran}
\end{thebibliography}
\end{document}